\documentstyle[epsf,amssymb]{elsart}

\newcommand{\rr   }   { {\bf r} }
\newcommand{\dr  }   { {\rm d}\rr\ }
\newcommand{\D   }[1]{ {\cal D}#1\ }
\newcommand{\PD  }[2]{\frac{\partial#1}{\partial#2}}
\begin{document}
\setlength{\baselineskip}{14pt}

\begin{frontmatter}

\title{Adsorption of Large Ions from an Electrolyte Solution:
A Modified Poisson--Boltzmann Equation \thanksref{submitted}}

\author{I. Borukhov \thanksref{UCLA}}

\address{Unit\'e mixte de Recherche CNRS/Elf Atochem (UMR 167), \\
95 rue Danton, B.P. 108, 92303 Levallois-Perret Cedex, France
}

\author{D. Andelman}

\address{School of Physics and Astronomy, \\
Raymond and Beverly Sackler Faculty of Exact Sciences, \\
Tel Aviv University, 69978 Ramat Aviv, Israel}

\author{H. Orland}

\address{Service de Physique Th\'eorique, CE-Saclay,
     91191 Gif-sur-Yvette, Cedex, France}

\thanks[submitted]{Submitted to Electrochimica Acta}

\thanks[UCLA]{Address from Jan. 2000:
Dept. of Chemistry, UCLA, Los Angeles CA 90024, USA}

\begin{abstract}
    The behavior of electrolyte solutions close to
a charged surface is studied theoretically. A modified
Poisson--Boltzmann equation which takes into account the volume
excluded by the ions in addition to the electrostatic interactions
is presented. In a formal lattice gas formalism the modified
Poisson--Boltzmann equation can be obtained from a mean--field
approximation of the partition function. In an alternative
phenomenological approach, the same equation can be derived by
including the entropy of the solvent molecules in the free energy.
In order to visualize the effect of the steric repulsion, a simple
case of a single, highly charged, flat surface is discussed. This
situation resembles recent adsorption experiments
of large ions onto a charged monolayer.
A simple criterion for the importance of the steric effects is
expressed in terms of the surface charge density and the size of
the ions. It is shown that when these effects are important a
saturated layer is formed near the surface. A modified Grahame
equation relating the ion concentration at the surface to the
surface charge density is obtained.
\end{abstract}


\end{frontmatter}


\section{Introduction}

The Poisson--Boltzmann equation is the main tool for studying the
behavior of ionic solutions
\cite{IsraelachviliBook,Adamson,DLVO,AndelmanPB}. Its main
advantages are its simplicity, which allows for analytical
solutions in simple cases, and its surprisingly good agreement
with experiments.
Aqueous solutions of small and macro-ions are of interest from an
industrial point of view ({\it e.g.,} colloidal suspensions
\cite{Hunter}) and as an important component in biological systems
({\it e.g.,} DNA, charged membranes). Therefore, the
Poisson--Boltzmann approach was  applied to many situations. Of
particular interest are: (i) adsorption of ions to flat surfaces
\cite{Henderson,Mirkin}; (ii) ion distribution around a charge
cylinder and the so-called {\em Manning condensation}
\cite{Katchalsky,Manning,Belloni,Netz,Holm}; and (iii) ion
distribution around a charged sphere and the so-called {\em charge
renormalization} \cite{Belloni,Alexander}.

In its simpler form, the linearized Poisson-Boltzmann equation
leads to the Debye--H\"uckel expression, thus providing a simple
description of screening effects in terms of the Debye--H\"uckel
screening length \cite{Debye}.
The success of the Poisson--Boltzmann approach
is quite impressive in view of the
various approximations which are included in its derivation:
it is a mean field approach which
totally neglects correlations and all specific (non-electrostatic)
interactions between the ions including the ionic finite size.

Nevertheless, despite its success in describing a wide range of
systems, it has been known for a long time to have
some limitations in several cases:
(i) the phase transition of electrolyte solutions \cite{Fisher},
(ii) the adsorption of charged ions to highly charged surfaces
\cite{Eigen,Iglic,Rondelez,prl},
and (iii) the attractive interactions which can be observed
between equally charged surfaces in the presence of asymmetric
electrolytes \cite{Kjellander,Kekicheff}.
Consequently, there have been numerous attempts to improve upon the
standard Poisson-Boltzmann approach
\cite{HNC,MSA,Marcelo,Hansen,Blankschtein,NetzOrland,Burak}.

\begin{figure}[tbh]
  \epsfxsize=8cm
  \centerline{\vbox{ \epsffile{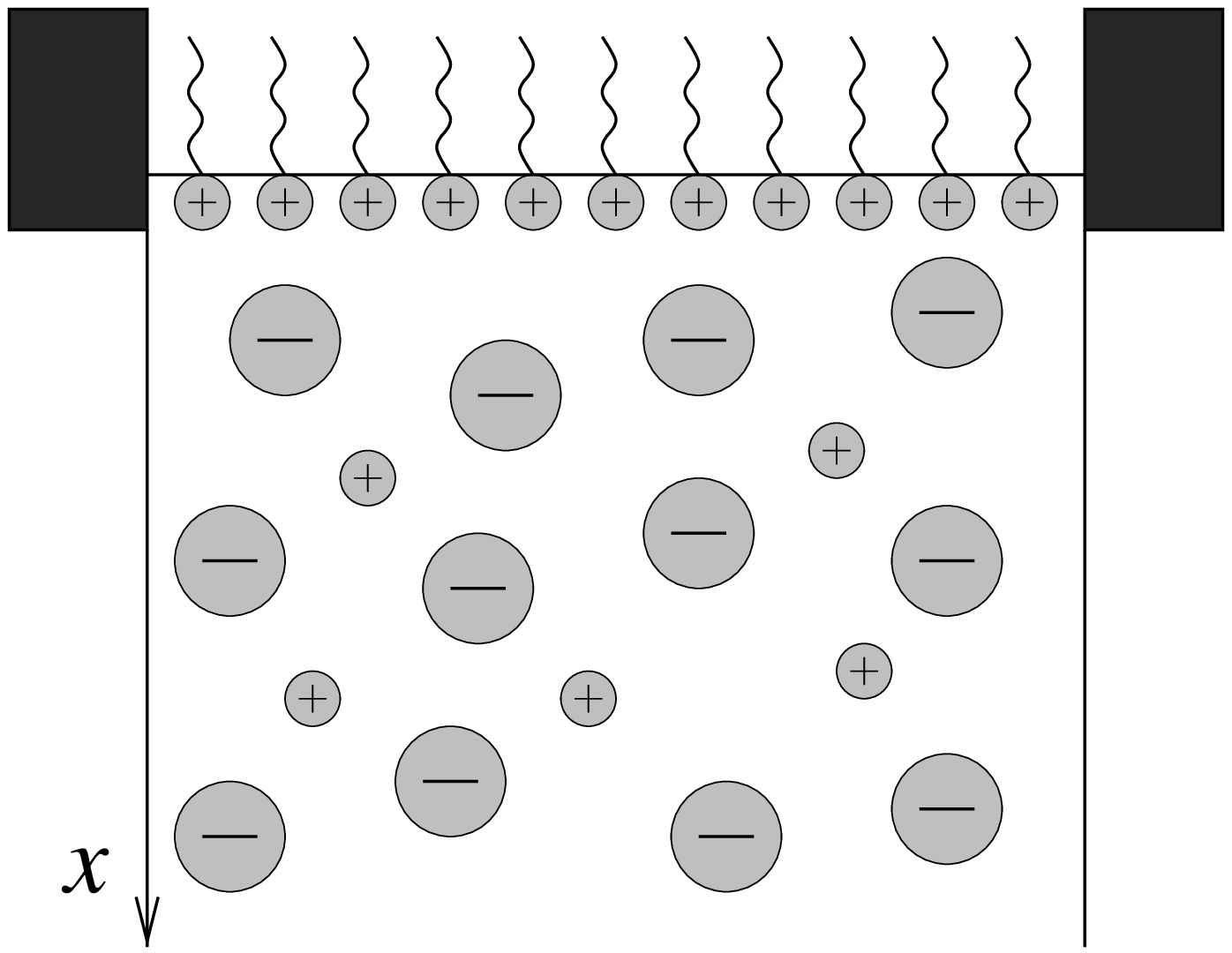}}}
 \caption[Adsorption of large ions to a charged monolayer]{
Schematic view of the adsorption of large ions to a charged
monolayer \cite{Rondelez}. The surface charge is carried by
amphiphilic molecules which are confined to the air/water
interface. The surface charge density can be varied continuously
by changing the area per amphiphilic molecule.}
\label{fig:Rondelez}
\end{figure}

  In this study we focus on the second case, where highly
charged surfaces attract a large amount of free ions from the
solution. At high ion densities achieved close to the
surface, short range ion--ion interactions become comparable
to the Coulomb interaction and they can no longer be neglected.
In particular, the ion density is bounded by the maximum
value which is obtained when the ions are closely packed.
 Recently, Cuvillier et al. \cite{Rondelez} have
provided an experimental setup which clearly demonstrates this
situation. In their experiments large tungstic acid ions (diameter
of about 10\AA) were adsorbed onto a charged monolayer spread at
the air/water interface (Fig.~\ref{fig:Rondelez}). The insoluble
(Langmuir) monolayer consists of charged amphiphilic molecules
having two moieties: a charged head group favoring the water side
of the interface, and a hydrocarbon tail favoring the air side.
The surface charge density can be controlled continuously by
changing the monolayer density through a lateral surface pressure.
In the experiments, a large discrepancy was found between the
measured ion concentration near the surface and the high values
anticipated by the Poisson--Boltzmann approach.

  Our aim in this study is to include the finite size of the ions
in the Poisson--Boltzmann approach and study how the ion
distribution close to charged surfaces is affected.
The standard way of including the finite size of the ions
in the Poisson--Boltzmann approach is to define a
narrow layer close to the surface as impenetrable
to the ions. This layer is usually referred to as the
{\em Stern layer} \cite{Stern}
and its width is equal to the ion radius.
Outside this layer the regular Poisson--Boltzmann
equation is implemented.
In our approach, a modified Poisson--Boltzmann equation is derived
where the steric forces lead to saturation of the ion
density at high potentials.
This way, the width of the saturated layer
depends also on the surface charge and is not limited a-priori
to only one counterion layer.

In the next section the modified PB equation is derived
for different combinations of ion valencies. Two derivations
are presented: first, a systematic path integral approach
where the ions are put on a discrete lattice,
and second, a phenomenological free energy formulation
where the excluded volume effect is added through the
entropy of the solvent.
This equation is then implemented in Sec. 3 to study the adsorption
of large ions to flat surfaces.

\section{The Modified Poisson--Boltzmann Equation}

\subsection{Lattice Gas Formulation}

  Consider an aqueous solution of charged ions.
For simplicity we will assume that both co-ions and counter-ions
have the same size $a$. This assumption can be justified when all
the surface charges are of the same sign, since only counterions
are then attracted to the surface and reach high charge densities.
  Another simplification is that we do not distinguish between
the counterions which dissociate from the charged surfaces and the
ones  originating from the added salt.

  For the valencies of the ions we will consider three cases:
(i) a symmetric $z$:$z$ electrolyte, (ii) an asymmetric $1$:$z$
electrolyte, and (iii) $z$-valent counterions without additional
salt. The different cases will be used to study the application of
the modified equation in different physical systems.

\begin{figure}[tbh]
  \epsfxsize=8cm
  \centerline{\vbox{ \epsffile{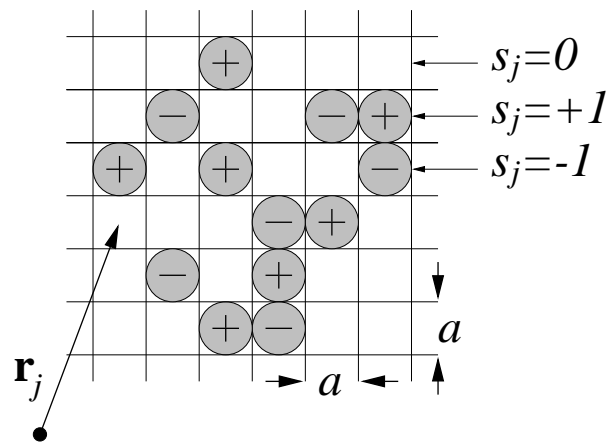} } }
\caption[Electrolyte lattice model] {
Schematic view of an electrolyte on a lattice model. The lattice
cells are located at $\rr_j$ and assigned a spin-like variable
$s_j=0,\pm 1$.}
\label{fig:lattice}
\end{figure}

  For a symmetric $z$:$z$ electrolyte the solution contains
two charge carriers, with charges equal to $\pm ze$. In order to
derive the free energy we will use a discrete lattice gas
formulation. In this approach, the charge carriers are placed on a
three dimensional cubic lattice where the dimensions of a single
cell are $a\times a\times a$ (Fig.~\ref{fig:lattice}).
Thus, by dividing space into discrete cells (lattice sites) and
limiting the occupation of each cell to a single ion we introduce
a short range repulsion between the ions. The size of a cell
represents the volume of an ion up to a numerical prefactor.

In order to describe the occupation of cells by ions we assign to
each cell $j$, which is located at $\rr_j$, a spin-like variable
$s_j$. This variable can have one of three values: $s_j=0$ if the
cell is empty (occupied by a water molecule), and $s_j=\pm 1$
according to the sign of the ion that occupies the cell.
The partition function of the system can now be written in the
form \cite{Coalson}
\begin{eqnarray}
   Z = \sum_{s_j=0,\pm 1}
   \exp\left(
     -{\beta\over 2} z^2e^2 \sum_{j,j'} s_j v_c(\rr_j-\rr_{j'}) s_{j'}
        + \sum_j \beta \mu_j s_j^2 \right)
\end{eqnarray}
  The first term in the exponent is the electrostatic energy, where
$v_c(\rr) = 1/\varepsilon|\rr|$ is the Coulomb interaction. The
second term is the chemical potential term where $\mu_j=\mu_+$ for
the positive ions ($s_j=+1$) and $\mu_j=\mu_-F$ for the negative
ions ($s_j=-1$).

  The charge density operator can be expressed in terms of the
spin variables as
\begin{eqnarray}
  \hat\rho_c(\rr) = \sum_j z e s_j \delta(\rr-\rr_j)
\end{eqnarray}
It is then useful to introduce the density field $\rho_c(\rr)$ and
the conjugate field $\varphi_c(\rr)$ through the identity
\begin{eqnarray}
  1 &=& \int\D{\rho_c}\delta\biggr[\rho_c(\rr) - \hat\rho_c(\rr)\biggl]
\nonumber\\
    &=& \int\D{\rho_c}\D{\varphi_c}
   \exp\biggl(
        i\beta \int\dr\rho_c\varphi_c
       -i\beta ze \sum_j s_j\varphi_c(\rr_j)
      \biggr)
\end{eqnarray}
where $\int\D{\rho_c}$ is a functional integral over the values of
$\rho_c$ at all space points $\rr$. It can viewed as the continuum
limit of multiple integrals over the values of $\rho_c$ at
different points in space (see, {\it e.g.,} ref. \cite{Wiegel}):
\begin{equation}\label{Df}
 \prod_j \int {\rm d}\rho_c(\rr_j) \rightarrow \int\D{\rho_c}
\end{equation}

Using the above identity is equivalent to the
Hubbard--Stratonovitch transformation (see, {\it e.g.,} ref.
\cite{OrlandNegele})
and leads to

\begin{eqnarray} \label{Zrhocphic}
   Z &=& \int \D{\rho_c}\D{\varphi_c}
   \exp\biggl(
    -{\beta\over2} \int\dr{\rm d}{\bf r}'\,\rho_c(\rr) v_c(\rr-\rr') \rho_c(\rr')
   \nonumber \\ &&~~~~~~~~~~~~~~~~~~~~~~~
    + i\beta \int\dr\rho_c(\rr)\varphi_c(\rr)
    \biggr)
   \nonumber \\ && ~~~~~~~~~~~~~
   \times
   \sum_{s_j=0,\pm 1}
      \exp\biggl(
        -i\beta ze \sum_j s_j\varphi_c(\rr_j)
        + \sum_j \beta \mu_j s_j^2 \biggr)
\end{eqnarray}
Recall that $\rr_j$ are the discrete coordinates of the lattice
sites while $\rr$,$\rr'$ are continuous spatial coordinates. It is
now possible to trace over the allowed values of the spin-like
variables, $s_j=0,\pm 1$ (last line in eq.~\ref{Zrhocphic}). In
the continuum limit, where physical properties vary on length
scales much larger than the size of a single site, the sum over
the lattice sites can be replaced by a continuous integral over
space and the partition function simplifies to
\begin{eqnarray}
   Z &=& \int \D{\rho_c}\D{\varphi_c}
   \exp\biggl(
    - {\beta\over2} \int\dr{\rm d}{\bf r}'\,\rho_c(\rr) v_c(\rr-\rr') \rho_c(\rr')
   \nonumber \\ &&~~~~~~~~~~
    + i\beta \int\dr\rho_c(\rr)\varphi_c(\rr)
   \nonumber \\ &&~~~~~~~~~~
   + {1\over a^3}\int\dr
   \ln\bigl\{ 1+\e^{\beta\mu_+-iz\beta e\varphi_c(\rr)}
               +\e^{\beta\mu_-+iz\beta e\varphi_c(\rr)}
   \bigr\} \biggr)
\end{eqnarray}
Since the exponential is quadratic in $\rho_c$, its functional
integral can be performed.

\begin{eqnarray} \label{Zpsi}
   Z = \int \D{\varphi_c}
   \exp\biggl(&-&{\beta\varepsilon\over8\pi} \int\dr |\nabla\varphi_c|^2
    \nonumber\\
   &+& {1\over a^3}\int\dr
   \ln\bigl\{ 1+\e^{\beta\mu_+-iz\beta e\varphi_c(\rr)}
               +\e^{\beta\mu_-+iz\beta e\varphi_c(\rr)}
   \bigr\} \biggr)
\end{eqnarray}

The chemical potentials $\mu_\pm$ are related to the total number
of positive and negative ions in the solutions through
\begin{eqnarray}\label{Npm}
   N_\pm = {1\over Z}\PD{Z}{(\beta\mu_\pm)}
     = \left\langle {1\over a^3}\int\dr
     {\e^{\beta\mu_\pm \mp iz\beta e\varphi_c(\rr)} \over
      1+\e^{\beta\mu_+-iz\beta e\varphi_c(\rr)}
               +\e^{\beta\mu_-+iz\beta e\varphi_c(\rr)}}
    \right\rangle
\end{eqnarray}
Where $\langle\cal{O}\rangle$ denotes the grand canonical average
of the operator $\cal{O}$.

In the bulk, the total number of positive and negative ions is
equal, $N_+=N_-=N/2$. It is useful to define the volume fraction
occupied by both the co- and counter-ions as $\phi_0=Na^3/V = 2c_b
a^3$ where $V$ is the total volume and $c_b$ is the bulk
concentration of the electrolyte. In the thermodynamic limit
$N,V\to\infty$ while $c_b$ and $\phi_0$ remain finite. Using
eq.~\ref{Npm} the chemical potentials can be expressed in terms of
$\phi_0$:
\begin{eqnarray}
   \e^{\beta\mu_+} = \e^{\beta\mu_-}
   = {1\over 2}{\phi_0\over 1-\phi_0}
\end{eqnarray}

  In the mean field approximation, the partition function is
approximated by the value of the functional integral at its saddle
point $\psi(\rr)\equiv i\varphi_c$. The free energy of the system
is then given by
\begin{eqnarray}
   {F\over k_BT} &=& - \ln{Z} \nonumber\\
     &=& -{\beta\varepsilon\over8\pi} \int\dr |\nabla\psi|^2
       - {1\over a^3}\int\dr
   \ln\left\{ 1+{\phi_0\over1-\phi_0}\cosh\left[\beta ze\psi(\rr)\right]
     \right\}
\end{eqnarray}
where $\psi(\rr)$ satisfies the {\em modified Poisson--Boltzmann
equation} for a symmetric $z$:$z$ electrolyte \cite{prl}:
\begin{eqnarray}
   \nabla^2\psi = {8\pi ze \over\varepsilon}
   {c_b \sinh(z\beta e\psi) \over
    1 - \phi_0 + \phi_0 \cosh(z\beta e\psi)}
   \label{PBztoz}
\end{eqnarray}

  In the zero size limit, $a\to 0$,
(namely, $\phi_0\to 0$ while $c_b$ remains fixed) the above
equation reduces to the regular Poisson--Boltzmann equation:
\begin{eqnarray}
   \nabla^2\psi = {8\pi ze \over\varepsilon} c_b\sinh(z\beta e\psi)
   \label{PBsalt2}
\end{eqnarray}

  For an asymmetric 1:$z$ electrolyte the derivation is very similar
and the modified PB equation is:
\begin{eqnarray}
   \nabla^2\psi = {4\pi z e c_b \over\varepsilon}
   {\e^{z\beta e\psi} - \e^{-\beta e\psi} \over
    1 - \phi_0 + \phi_0 (\e^{z\beta e\psi}
            + z\e^{-\beta e\psi})/(z+1)}
  \label{PB1toz}
\end{eqnarray}
  where $\phi_0 = (z+1) a^3 c_b$ is the combined bulk
volume fraction of the positive and negative ions.

Finally, if the solution is salt free and contains only negative
counterions of valency $-|z|$, the modified PB equation becomes
\begin{eqnarray}
   \nabla^2\psi = {4\pi z e c_0 \over\varepsilon}
   {\e^{z\beta e\psi} \over
    1 - \phi_0 + \phi_0 \e^{z\beta e\psi}}
  \label{PB0toz}
\end{eqnarray}
  where $\phi_0 = a^3 c_0$ is the volume fraction at an arbitrary
reference point $\rr_0$ where $\psi(\rr_0)=0$ and $c(\rr_0)=c_0$.
Note that the reference point of zero potential does not lie at
infinity. The salt-free system contains only the counterions which
neutralize the surface charges. Since the surface is taken to be
infinite in its size, the potential does not go to zero as
$x\rightarrow \infty$, but it diverges to $-\infty$. Physically
this divergence is not a problem because the electric field and
counterion density tend to zero at large distances.

\subsection{Phenomenological Free Energy Derivation}

  The modified PB equation (eqs.~\ref{PBztoz},~\ref{PB1toz}
and \ref{PB0toz}) can also be derived from a phenomenological free
energy \cite{prl}. Let us consider again the symmetric $z$:$z$
case. This is done by expressing the free energy of the system
$F=U_{\rm el} - TS$ in terms of the local electrostatic potential
$\psi(\rr)$ and the ion concentrations $c^\pm(\rr)$. The
electrostatic contribution is
\begin{eqnarray}
  U_{\rm el} = \int\dr \biggl[-{\varepsilon\over 8\pi}|\nabla\psi|^2
    + zec^+\psi - zec^-\psi
      -\mu_+c^+  -\mu_-c^-
               \biggr]
\end{eqnarray}
  The first term is the self energy of the electric field
and the next two terms are the electrostatic energies of the ions.
The last two terms couple the system to a bulk reservoir, where
$\mu_\pm$ are the chemical potentials of the ions.

The entropic contribution is
\begin{eqnarray}
  -TS = {k_BT \over a^3}  \int\dr
         \biggl[ &&{c^+a^3}\ln\bigr(c^+a^3\bigr) +
                 {c^-a^3}\ln\bigl(c^-a^3\bigr)  \nonumber\\
      &+&  \bigl(1-c^+a^3-c^-a^3\bigr)
           \ln\bigl(1-c^+a^3-c^-a^3\bigr) \biggr]
\end{eqnarray}
The first two terms represent the translational entropy of the
positive and negative ions, whereas the last term is the entropy
of the solvent molecules. It is this last term that is responsible
for the modification of the PB equation. Note that it is also
possible to include additional short-range (non-electrostatic)
interaction terms in the MPB free energy \cite{Burak}, but we will
not consider them here.

Minimizing the total free energy with respect to $\psi$ and
$c^\pm$ yields the Poisson equation
\begin{eqnarray}
   \nabla^2\psi = - {4\pi\over \varepsilon}
   \left[zec^+(\rr) - zec^-(\rr) \right]
\end{eqnarray}
where the ion concentrations are given by
\begin{eqnarray}
 c^\pm = {c_b \e^{\mp \beta z e\psi} \over
     1 - \phi_0 + \phi_0 \cosh\left(\beta ze\psi\right)}
\label{cpm}
\end{eqnarray}
Here, as before, $\phi_0=2c_b a^3$ denotes the bulk volume
fraction of the small ions. Combining the above two expressions
recovers the modified PB equation (eq.~\ref{PBztoz}). The same
approach can also be applied in the derivation of
eqs.~\ref{PB1toz}, \ref{PB0toz} \cite{prl}.
A similar expression was suggested in the 50's by Eigen
\cite{Eigen} and more recently by Kralj-Igli\v{c} and Igli\v{c}
\cite{Iglic}. Similar ionic distributions can be also obtained for
solid electrolytes \cite{Gurevich,Kornyshev}.

This approach deviates significantly from the original PB equation
for large electrostatic potentials $|\beta e \psi|\gg 1$. In
particular, the ionic concentration is
 unbound in the standard PB approach, whereas here
it is always bound by $1/a^3$ (``close packing'') as can be seen
from eqs.~\ref{PBztoz}, \ref{PB1toz} and \ref{PB0toz}. This effect
is important close to strongly charged surfaces immersed in an
electrolyte solution.

Note that for high positive potentials, $\beta e \psi \gg 1$, the
contribution of the positive ions is negligible and the negative
ion concentration follows a distribution reminiscent of the
Fermi-Dirac distribution
\begin{eqnarray}
c^-(\rr) \to {1 \over a^3} ~\frac{1}{1+\e^{-\beta(z e\psi+\mu)}}
  \label{FermiDirac}
\end{eqnarray}
where the excluded volume interaction plays the role of the Pauli
exclusion principle, and $\mu=\mu_-$ is the chemical potential of
the negative ions.

The effect of the additional entropy term and the fact that it
limits the concentration of ions near the surface can be
demonstrated by the following simple argument: consider a nearly
saturated region near a highly charged (positive) surface where
$c^-a^3\to 1$. On one hand, the system can gain entropy by pushing
solvent molecules in between the positive ions. On the other hand,
diluting the saturated layer costs electrostatic energy.
Altogether, the excess free energy per unit volume of diluting the
saturated layer with a volume fraction $\eta=1-c^-a^3$ of solvent
molecules is:
\begin{eqnarray}
  \Delta f \simeq {k_BT\over a^3}\eta\left(\ln{\eta}-1\right)
    + {ze|\psi_s|\over a^3} \eta
\end{eqnarray}
$\psi_s$ is the electric potential near the surface. The balance
between these two terms results in an optimal non-zero dilution
equal to
\begin{eqnarray}
  \eta^* \simeq \e^{-z\beta e|\psi_s|}
\end{eqnarray}
and the free energy gain due to this dilution is
\begin{eqnarray}
  \Delta f \simeq - {k_BT\over a^3} \eta^*
\end{eqnarray}
We conclude that entropy will always drive solvent molecules
into the saturated layer, although at high surface charge
densities their amount becomes exponentially small.

\section{Adsorption to a Flat Surface}

The modified PB equation can help to interpret the adsorption
experiments of Cuvillier et al. \cite{Rondelez}.
In their experiments large (diameter of about $10$\AA) negative
multivalent ($z=3$ or $z=4$) ions were adsorbed onto a positively
charged Langmuir monolayer (Fig.~\ref{fig:Rondelez}). The large
polyanions such as phosphotungstic acid (H$_3$PW$_{12}$O$_{40}$)
were dissolved in an aqueous subphase and attracted to a cationic
Langmuir monolayer such as a fatty amine surfactant
(C$_{20}$H$_{41}$-NH$_2$), spread at the water/air interface.

Consider, therefore, a solution containing an asymmetric (1:$z$)
electrolyte in contact with a single planar surface of charge
density $\sigma>0$. Assuming that the system is homogeneous in the
lateral directions, the modified PB equation (eq.~\ref{PB1toz})
reads:
\begin{eqnarray}
   y''(x) = 4\pi l_B z c_b
   {\e^{z y} - \e^{-y} \over
    1 - \phi_0 + \phi_0 (\e^{z y} + z\e^{-y})/(z+1)}
  \label{PBflat}
\end{eqnarray}
where $y(x)=\beta e\psi(x)$ is the reduced electrostatic potential
as function of  the distance $x$ from the surface. and
$l_B=e^2/\varepsilon k_BT$ is the Bjerrum length. $l_B=7$\AA\ for
an aqueous solution at room temperature.
Equation~\ref{PBflat} can be solved numerically yielding the
electrostatic potential and ion concentration profiles
as a function of the distance from the surface.

\begin{figure}[tbh]
  \epsfxsize=8cm
\centerline{\vbox{\epsffile{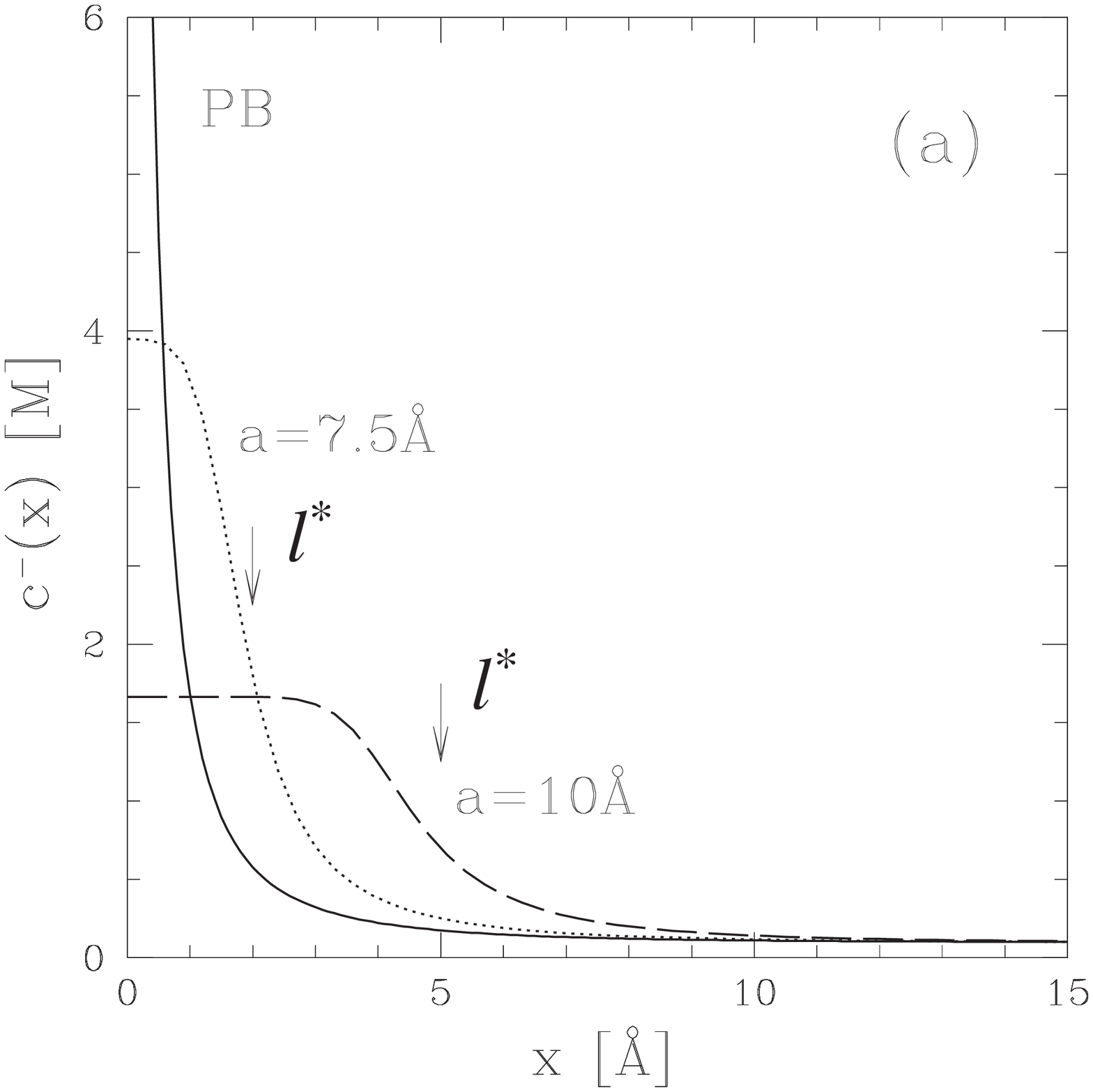} }
  \epsfxsize=8cm
            \vbox{ \epsffile{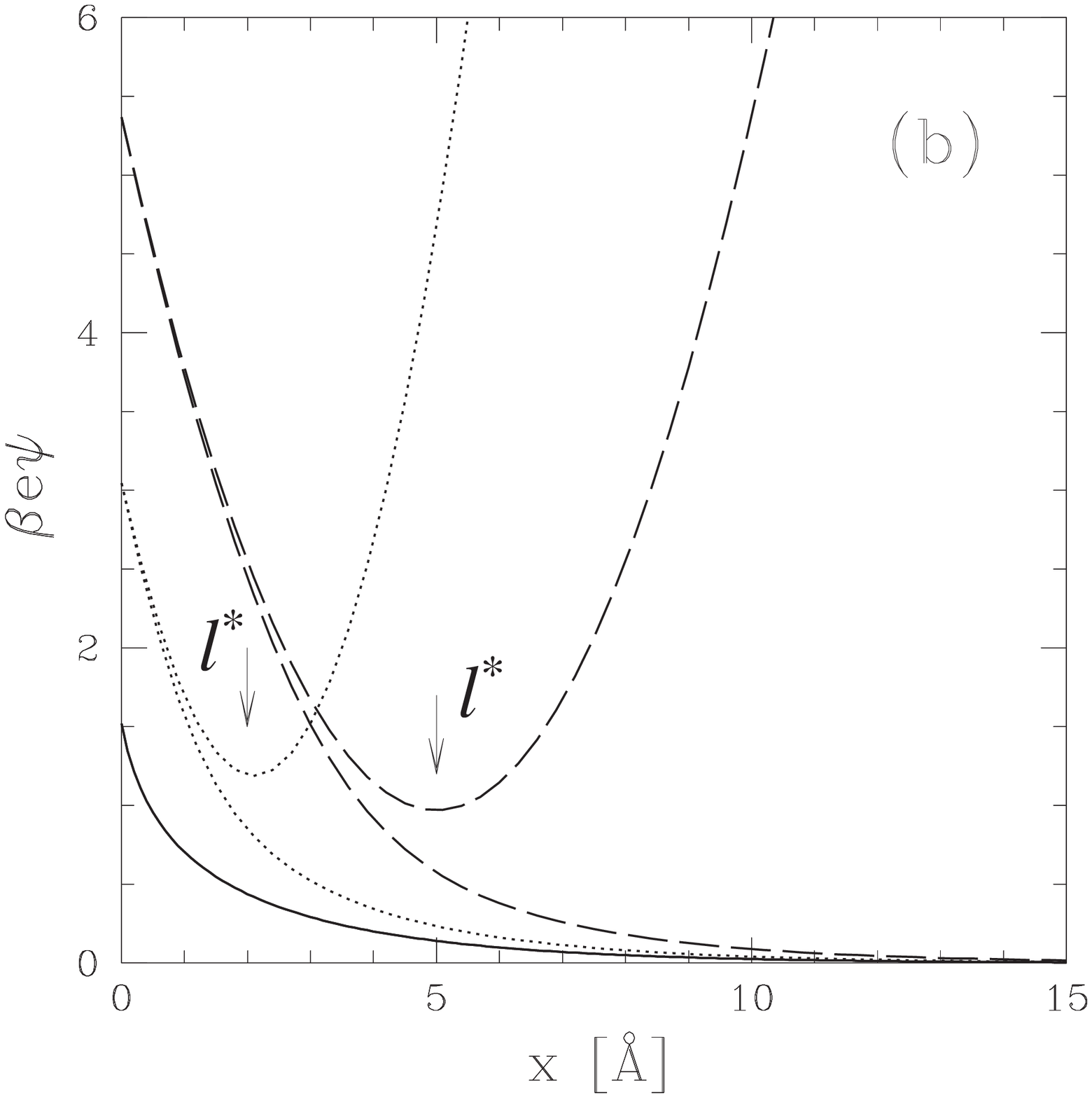} } }
\caption[Electrolyte adsorption profiles near a flat surface] {
(a) Concentration profiles of negative
multivalent ions $c^-(x)$ near a positively charged surface as
obtained from the numerical solution of eq.~\ref{PBflat} for two
different ion size $a=7.5$\AA\ (dotted line)  and $a=10$\AA\
(dashed line). The saturated layer width $l^*\simeq2$\AA\ and
$5$\AA, respectively, is indicated by small arrows. The solid line
represents the concentration profile of the standard PB equation.
 (b) Calculated electrostatic potential profiles near
the surface plotted together with the parabolic approximation
(eqs.~\ref{psi-strong}, \ref{psi-s}). The dotted, dashed and solid
lines are as in (a). The bulk concentration is $c_b=0.1$M for a
1:$z$ electrolyte with $z$=4. The  surface charge density $\sigma$
is taken as one electron charge per 50\AA$^2$. The aqueous
solution with $\varepsilon=80$ is at room temperature so that the
Bjerrum length is $l_B=7$\AA.}
\label{fig:profiles}
\end{figure}

Typical results are presented in Fig.~\ref{fig:profiles} for
various values of the ion size $a$ together with the solution of
the original ($a=0$) PB equation. In Fig.~\ref{fig:profiles}a the
negative ion concentrations are plotted and in
Fig.~\ref{fig:profiles}b the electric potential. Since at high
surface charge densities the positive ion concentration is small
near the surface, only the negative ion profiles are shown.

Clearly, the ionic concentration saturates to its maximal value in
the vicinity of the charged surface. This should be contrasted
with the original PB scheme which leads to extremely high and
unphysical values of  $c^-_s\equiv c^-(0)$, especially for
multivalent ions. In the saturated region, the ionic concentration
tends to $1/a^3$, leading to more pronounced deviations from PB
for larger ions. Note, that in contrast with the Stern layer
model, the width of the saturated layer (Fig.~\ref{fig:profiles}a)
is not strictly equal to $a$. As will be demonstrated below, it
depends not only on the parameter $a$ but also on the surface
charge density $\sigma$.

In the saturated layer the right--hand side of eq.~\ref{PB1toz}
becomes a constant, and the electrostatic potential is quadratic
\begin{eqnarray}
\label{psi-strong}
\psi(x) \simeq \psi_s - {4\pi\sigma\over\varepsilon}x
     + {2\pi ze\over\varepsilon a^3} x^2
\end{eqnarray}
where $\psi_s$ is the surface potential and the boundary condition
$\psi'|_s = -4\pi\sigma/\varepsilon$ is satisfied. As can be seen
in Fig.~\ref{fig:profiles}b, the parabolic curve is a good
approximation for $\psi(x)$ close to the surface. The width of the
saturated layer $l^*$ is not strictly equal to $a$. It can be
easily estimated from eq.~\ref{psi-strong} to be
\begin{eqnarray}
l^*\simeq a^3\sigma/z e
\label{lstar}
\end{eqnarray}
in qualitative agreement with Fig.~\ref{fig:profiles}a.

The surface potential $\psi_s$ can be calculated in a closed
analytical form using the first integral of eq.~\ref{PBflat}, with
a single assumption  that the positive ions (co-ions) density is
negligible at the surface
\begin{eqnarray}
  \psi_s ~\simeq~ {k_B T\over ze}\biggl\{
      \ln\Bigl[ \e^{\zeta_1} -\bigl(1-\phi_0 \bigr) \Bigr]
    - \ln\bigl(c_b a^3\bigr) \biggr\}
   ~\simeq ~ {k_B T\over ze}
          \biggl\{ \zeta_1 - \ln\bigl(c_b a^3\bigr) \biggr\}
  \label{psi-s}
\end{eqnarray}
   where
\begin{eqnarray}
  \label{zeta1}
  \zeta_1 \equiv  {2\pi a^3\sigma^2\over \varepsilon k_B T}
\end{eqnarray}
   The last approximation in eq.~\ref{psi-s} is valid only for
high values of $\zeta_1\gg1 $.
   The dimensionless parameter $\zeta_1$ is a measure of the
importance of excluded volume interactions. It can be written
as a ratio between the volume of a single ion and an electrostatic
volume
\begin{eqnarray}
    \zeta_1 = {a^3 \over \Sigma\lambda_{\rm GC}}
\end{eqnarray}
 where $\Sigma=e/|\sigma|$ is the area per surface charge,
and $\lambda_{\rm GC}=\Sigma/(2\pi l_B)\sim 1/|\sigma|$ is the
{\em Gouy--Chapman length} \cite{GC} characterizing the width of a
diffusive electrolyte layer near a charged surface in a salt-free
solution.

  Similarly, the concentration of negative ions at the surface
can be calculated leading to a modified Grahame Equation
\cite{IsraelachviliBook}
\begin{eqnarray}
  \label{cs}
  c^-_s \simeq {1\over a^3}\Bigl[1- (1-\phi_0)\e^{-\zeta_1}\Bigr]
\end{eqnarray}
In the above equation, as in eq.~\ref{psi-s}, the only additional
approximation is to neglect the concentration $c_s^+$ of the
positive (co-ions) close to the surface. At low surface charge
$\zeta_1\ll 1$, and the ion concentration reduces to the PB
results
\begin{eqnarray}
  \label{cs-pb}
  c^-_s = {2\pi\sigma^2 \over \varepsilon k_BT} + (1+z)c_b
\end{eqnarray}
but for high surface charge $\zeta_1 \gg 1$, the deviation from
the PB case is substantial.

\begin{figure}[tbh]
 \epsfxsize=8cm
 \centerline{\vbox{\epsffile{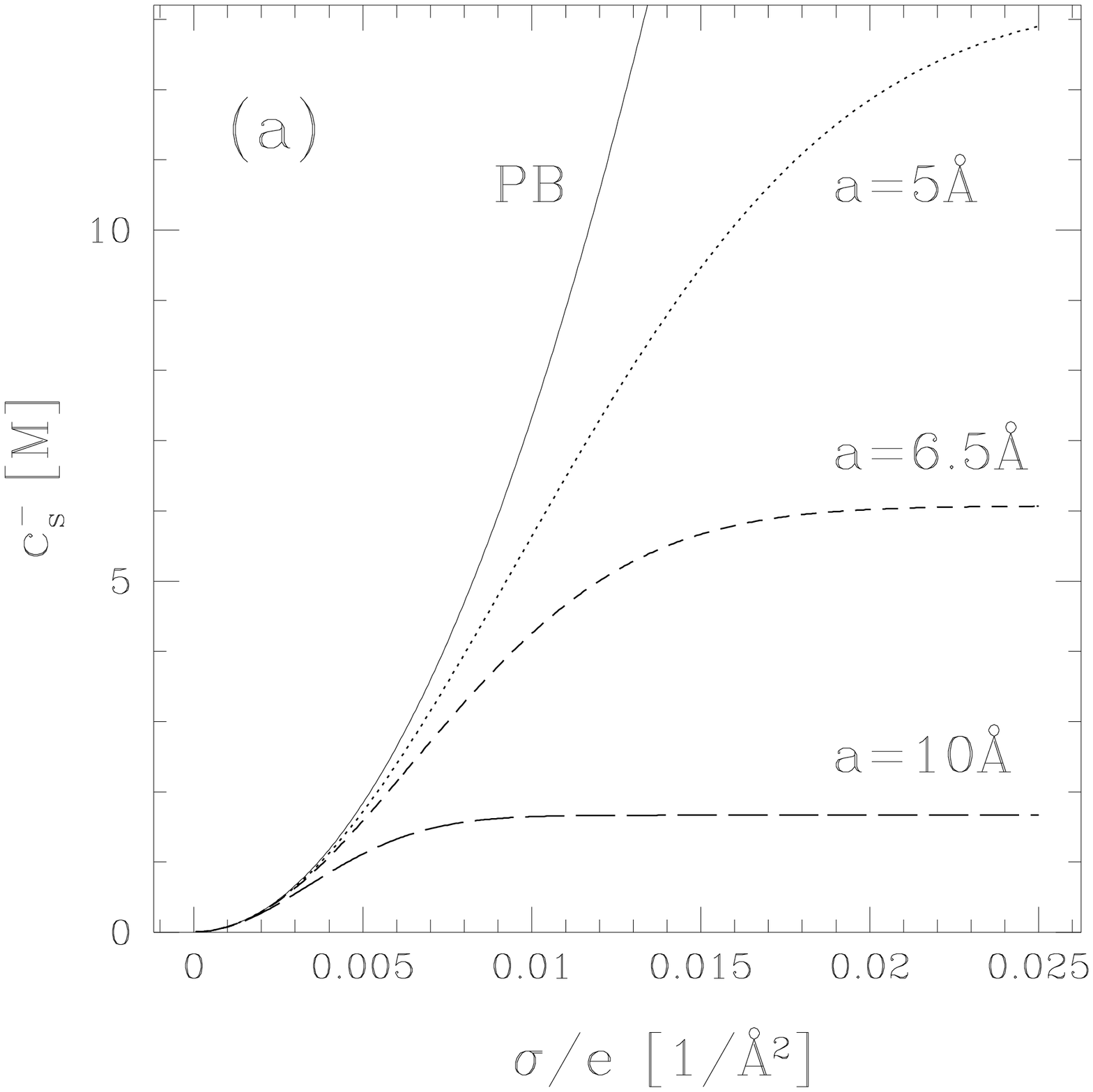} }
 \epsfxsize=8cm
             \vbox{\epsffile{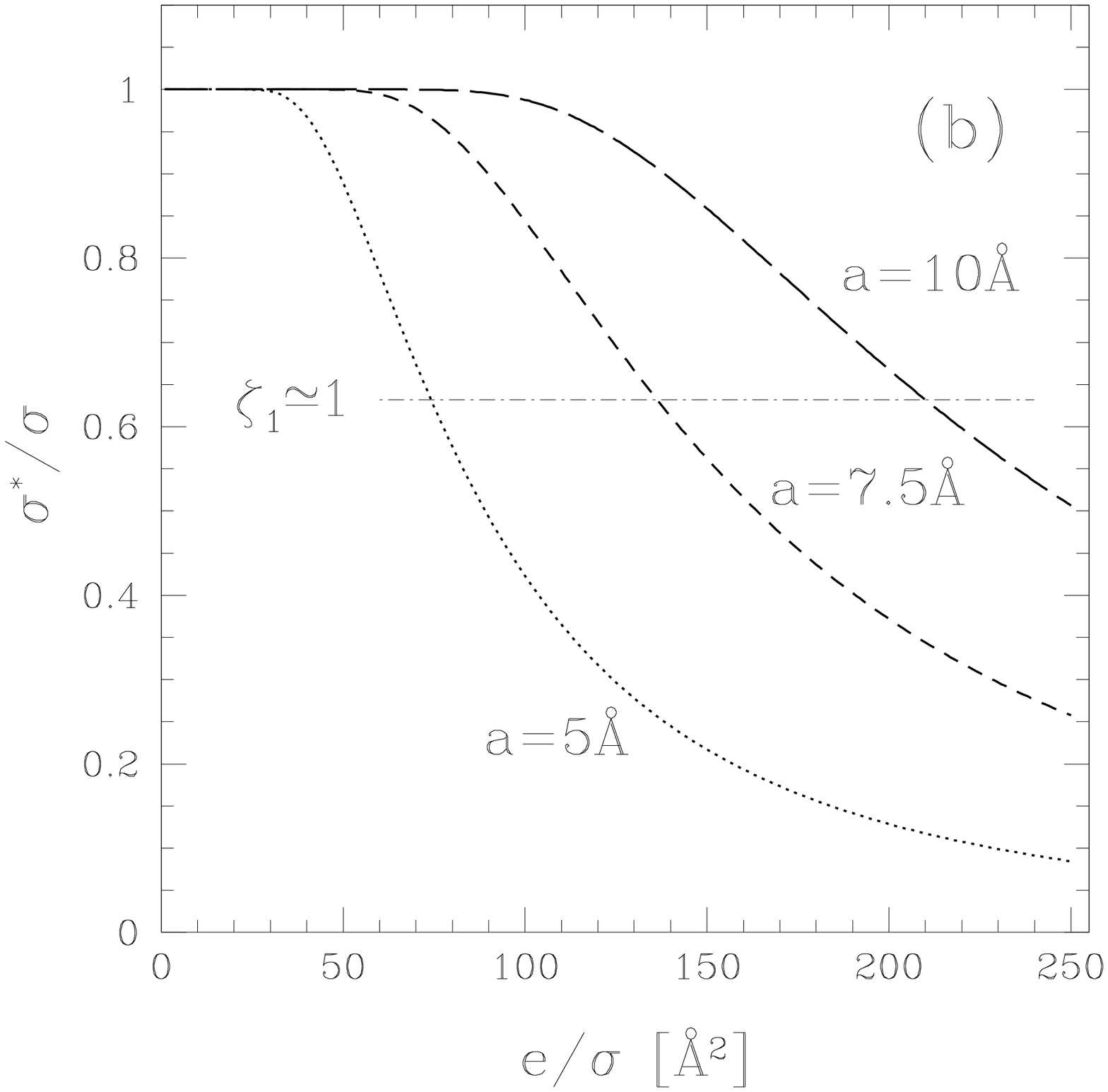} } }
\caption[Effect of surface charge on electrolyte adsorption] {
(a) Concentration of counterions at the
surface, $c_s^-=c^-(0)$, as a function of the surface charge
density $\sigma/e$ for different ions size, $a$. The PB
concentration is also plotted for comparison.
 (b) Ratio of the saturated layer charge density
and the surface charge density, $\sigma^*/\sigma=ze c^-_s
l^*/\sigma$, as a function of the specific surface area per unit
charge, $e/\sigma$, for different ion sizes. The 1:$z$ electrolyte
bulk concentration is $c_b=1m$M and the valency $z=4$. The
horizontal line indicates the crossover region $\zeta_1\simeq 1$
in low salt concentrations.}
\label{fig:surface_vals}
\end{figure}

 The surface concentration
of negative ions is depicted in Fig.~\ref{fig:surface_vals}a,
where $c^-_s$ is plotted as a function of the surface charge
density, $\sigma/e$. Equation~\ref{cs} is presented for three
different ions sizes and the original PB case (eq.~\ref{cs-pb}) is
shown as well for comparison. At low surface charges the four
curves are similar, but as the surface charge increases, the
curves deviate from each other. The large ions deviate first and
saturate to a lower surface concentration $c^-_s \to 1/a^3$.
The deviation point corresponds to $\zeta_1\simeq 1$. For
$a=10$\AA\ this gives $\sigma/e\simeq 0.005$\AA$^{-2}$ in
qualitative agreement with Fig.~\ref{fig:surface_vals}a.
It is interesting to note that the ionic concentration near the
surface at high surface charge density depends only weakly on the
bulk electrolyte concentration, $c_b$.

At high values of $\zeta_1$, the width of the saturated layer is
of the order of $l^*$ (eq.~\ref{lstar}) and the amount of charge
in the saturated layer can be estimated by
\begin{eqnarray}
   \sigma^* \simeq ze c^-_s l^*
            \simeq \left[ 1- (1-\phi_0)\e^{-\zeta_1} \right]\sigma
   \label{sigma_star}
\end{eqnarray}
In Fig.~\ref{fig:surface_vals}b the ratio between the ion charge
density of the saturated layer $\sigma^*$ and $\sigma$ is plotted
as function of the specific surface area per unit charge, as is
often measured in experiments.
At high surface charge densities (or equivalently, small surface
area per unit charge) the saturated layer plays a dominant role in
neutralizing the surface charge density. As the surface charge
density is lowered, the width of the saturated layer decreases
until it vanishes. This occurs when $\zeta _1$ is of order unity,
corresponding to $\sigma^*/\sigma\simeq 1-1/\e$ in the limit
$\phi_0\ll 1$. This crossover is indicated in
Fig.~\ref{fig:surface_vals}b as an horizontal line below which the
saturated layer is no longer well defined.
It should be noted that in our approach the counterions never
over-compensate the surface charges, namely, $\sigma^*/\sigma<1$.

In the experiments of Cuvillier et al. \cite{Rondelez} the
adsorbed ion density (per unit area) in the solution,
 $\sigma^*$, is measured by X-ray reflectivity.
It is then related to the surface charge density $\sigma$, which
is controlled by the Langmuir trough lateral pressure. The
experiments show very clearly the presence of the steric effects
for these large ions (of estimated size of 10\AA). As the surface
charge density increases, $\sigma^*/\sigma$ decreases in accord
with our findings (Fig.~\ref{fig:surface_vals}b) and in contrast
to the original PB approach.
The experiments also show some evidence that the amount of charge
in the adsorbed layer (per unit area) might exceed the original
surface charge density and lead to over-compensation of surface
charges \cite{Rondelez}. Our approach does not yield such an
effect. It is possible that additional ion-ion or ion-surface
interactions are responsible for this effect.

\begin{figure}[tbh]
 \epsfxsize=8cm
 \centerline{\vbox{\epsffile{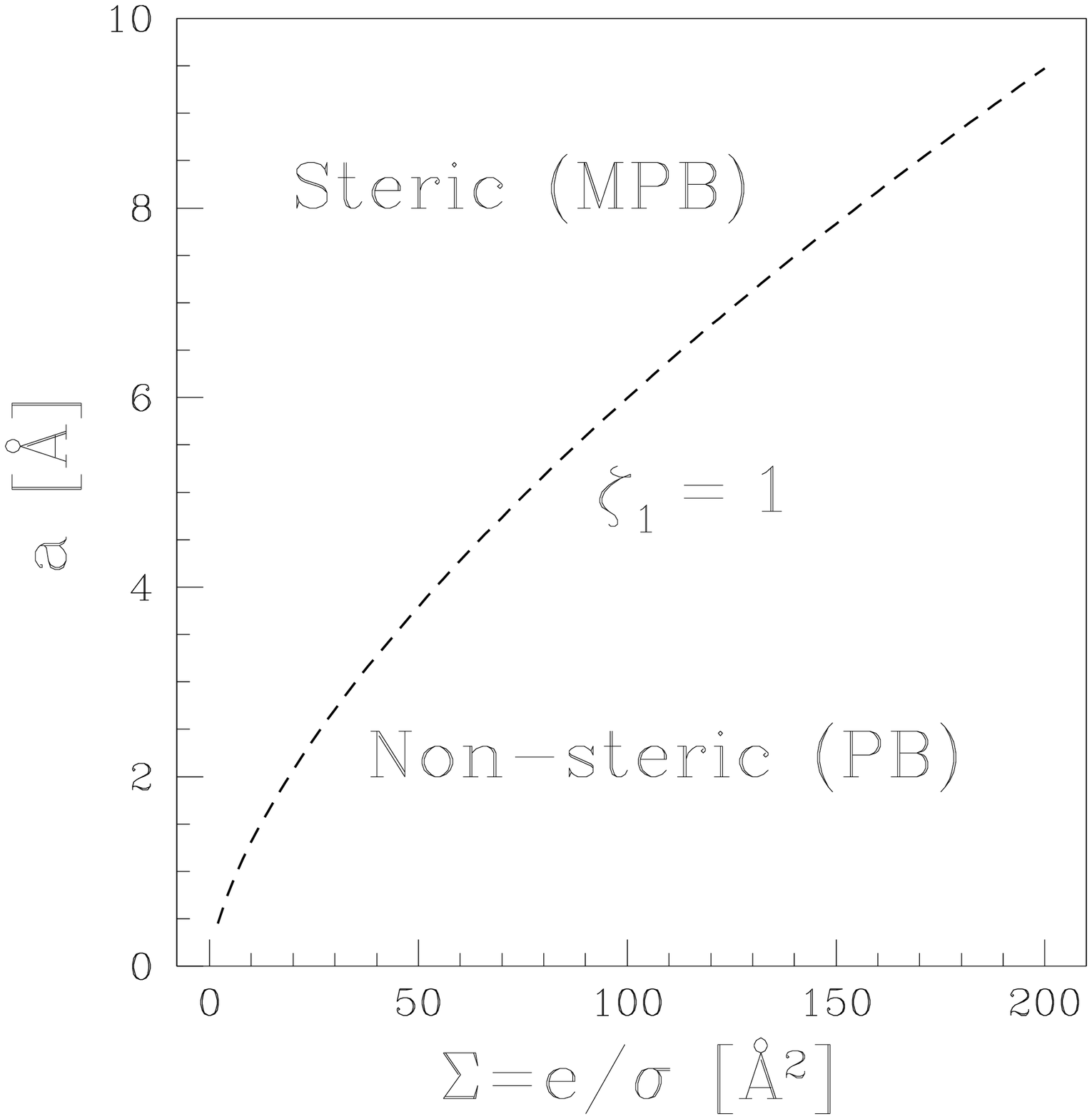} }  }
\caption[Large ion adsorption regimes in the planar case] {
Schematic diagram of the adsorption regimes
of the large ions to a charged surface. The dashed line separates
between the regime where steric effects are too weak to be
relevant and the regime where the short range ion-ion repulsion
becomes important. The horizontal axis is the area per unit charge
$e/\sigma$ in \AA$^2$ and the vertical axis is the ion size $a$.}
\label{fig:zeta1}
\end{figure}

The two different adsorption regimes are shown in
Fig.~\ref{fig:zeta1} where the dividing line $\zeta_1=1$ marks the
onset of steric effects. We have seen that at low charge densities
the parameter $\zeta_1\ll 1$ and our results coincide with those
of the original PB equation which does not take into account the
steric effects. On the other hand, when $\zeta_1\gg 1$,
corresponding to high surface charge densities or large ion sizes,
steric effects are, indeed, important. The concentration of
counterions can not exceed its maximal value of close packing and
this affects the surface potential and its dependence on the
surface charge.


\section{conclusions}

   In this work we have studied the effect of ion size on the
density profiles of counterions near a charged surface. For this
purpose, a modified Poisson-Boltzmann  (MPB) equation was derived
which includes the short range steric repulsion between ions in
addition to the electrostatic interactions. A formal lattice gas
derivation was presented as well as a phenomenological one. In the
former derivation the steric repulsion results from the maximal
occupancy of a single lattice site, while in the latter it stems
from the translational entropy of the ions and solvent molecules.
The two points of view are equivalent and result in the same MPB
equation.

   The MPB formalism was applied to a simple case of relatively
large counterions adsorbing to a strongly charged surface. This
model system corresponds to recent experiments \cite{Rondelez}
where multivalent large ions were adsorbed to a charged Langmuir
monolayer. The advantage of the monolayer setup is the possibility
to control the surface charge density via application of a surface
pressure at the water/air interface.

   We show that at high surface charge densities, in agreement
   with experiments, a saturated
counterion layer is formed close to the surface. This is due to
the fact that the ion concentration is limited by its maximal
value, namely, closed packing. This is in contrast with the
standard Poisson-Boltzmann formalism where the ion concentration
can be arbitrarily high.

The difference between the MPB and standard PB formalism can be
also expressed in terms of a modified Grahame equation relating
the concentration of counterions at the surface to the surface
charge density. This comparison introduces a dimensionless
parameter $\zeta_1=a^3/\Sigma\lambda_{\rm GC}$ where $a$ is the
ion size, $\Sigma$ the specific area per surface charge and
$\lambda_{\rm GC}=\Sigma/2\pi l_B$ the Gouy-Chapman length, $l_B$
being the Bjerrum length. As long as
$\zeta\lesssim 1$
the steric effects are weak. However, when
$\zeta\gtrsim 1$
(large $a$, small $\Sigma$) they cannot be neglected any more and
saturation appears in the adsorbed layer.

The quadratic corrections to mean field theory presented here can
be easily calculated. They correspond to the {\em Random Phase
Approximation} (RPA). In a homogeneous system, it can be shown
that the effective charge-charge interactions are of the
Debye-H\"uckel type with an unmodified screening length. In a
non-uniform system (e.g. in the presence of a charged surface) the
calculation is more involved. It requires knowledge of the
analytical solution of the corresponding mean field equation
\cite{NetzOrland2}.

Since the aim of this work was to identify and clarify the
consequences of a single effect, namely the ionic size, we did not
include additional contributions which might affect similar
systems. Among these are specific ion-surface interactions,
attractive ion-ion interactions and ion-ion correlations
\cite{HNC,MSA,Marcelo,Hansen,Blankschtein,NetzOrland,Burak}. The
latter are especially important for multivalent ions but their
inclusion usually leads to complex integral equations where the
underlying physics is not as transparent as in the MPB approach.

The MPB equation can be also applied to different geometries to
study the adsorption to curves surfaces \cite{pb-plus}
and the interactions
between charged surfaces in the presence of large ions.


\begin{ack}
We wish to thank N. Cuvillier and F. Rondelez for sharing with us
their experimental results prior to publication and for fruitful
discussions. We also benefited from discussions and correspondence
with H. Diamant, M. E. Fisher, C. Holm, A. A. Kornyshev, R. Netz,
P. Sens and M. Urbakh. Partial support from the US-Israel
Binational Foundation (BSF)  under grant No. 98-00429, and the
Israel Science Foundation founded by the Israel Academy of
Sciences and Humanities --- centers of Excellence Program is
gratefully acknowledged. One of us (HO) would like to thank the
Sackler Institute of Solid State Physics (Tel Aviv University) for
a travel grant. IB gratefully acknowledges the support of the
Chateaubriand postdoctoral fellowship.
\end{ack}


\vfill
\pagebreak
\section*{Figures}

\vspace{2cm}

\vspace{2cm}
\vfill
\pagebreak

\vspace{2cm}
\vfill

\vspace{2cm}

\vfill


\begin{thebibliography}{999}

\bibitem{IsraelachviliBook}
    J. N. Israelachvili,
   {\em Intermolecular and Surface Forces, 2nd ed.},
   Academic Press, London (1990).

\bibitem{Adamson}
    A. W. Adamson,
   {\em Physical Chemistry of Surfaces, 5th ed.},
   Wiley, New York (1990).

\bibitem{DLVO}
   B. V. Derjaguin and L. D. Landau,
   {\em Acta Phys. Chem. USSR} {\bf XIV}, 633 (1941);
    E. J. W. Verwey and  J. Th. G. Overbeek,
   {\em Theory of the Stability of Lyophobic Colloids},
   Elsevier, Amsterdam (1948).

\bibitem{AndelmanPB}
    D. Andelman,
    {\em Electrostatic Properties of Membranes: The Poisson--Boltzmann Theory"}
    chapter 12 in {\em Handbook of Biological Physics: Structure
    and Dynamics of Membranes},
    (Edited by R. Lipowsky and E. Sackmann), vol. 1B, pp. 603-642,
    Elsevier, Amsterdam (1995).

\bibitem{Hunter}
     R. J. Hunter,
     {\em Foundations of Colloid Science},
     Oxford University, New York (1989).

\bibitem{Henderson}
  D. Henderson,
  {\em Prog. Surf. Sci.} {\bf 13}, 197 (1983), 
  and references therein.

\bibitem{Mirkin}
   A. G. Volkov, D. W. Deamer, D. L. Tanelian and V. S. Mirkin,
  {\em Prog. Surf. Sci.} {\bf 53}, 1 (1997) 
  and references therein.

\bibitem{Katchalsky}
   R. M. Fuoss, A. Katchalsky and S. Lifson,
  {\em Proc. Natl. Acad. Sci. USA} {\bf 37}, 579 (1951). 

\bibitem{Manning}
    G. S. Manning,
  {\em J. Chem. Phys.} {\bf 51}, 924 (1969); 
  {\em ibid.} 934; 
  {\em ibid.} 3249. 

\bibitem{Belloni}
   L. Belloni,
  {\em Coll. Surf. A.} {\bf 140}, 227 (1998). 

\bibitem{Netz}
    R. R. Netz and J. F. Joanny,
  {\em Macromolecules} {\bf 31}, 5123 (1998). 

\bibitem{Holm}
    M. Deserno, C. Holm and S. May,
    {\em submitted to Macromolecules}.

\bibitem{Alexander}
  S. Alexander, P. M. Chaikin, P. Grant, G. J. Morales,
  P. Pincus and D. Hone,
  {\em J. Chem. Phys.} {\bf 80}, 5776 (1984). 

\bibitem{Debye}
    P. Debye and E. H\"uckel,
    {\em Physik} {\bf 24}, 185 (1923); 
    {\it ibid}, 305. 

\bibitem{Fisher}
    For a review see:
    M. E. Fisher,
    {\em J. Stat. Phys.} {\bf 75}, 1 (1994). 

\bibitem{Eigen}
   M. Eigen and E. Wicke,
   {\em J. Phys. Chem.} {\bf 58}, 702 (1954). 

\bibitem{Iglic}
   V. Kralj-Igli\v{c} and A. Igli\v{c},
   {\em Electrotechnical Rev. (Slovenia)} {\bf 61}, 127 (1994). 
   V. Kralj-Igli\v{c},
   {\em Electrotechnical Rev. (Slovenia)} {\bf 62}, 104 (1995); 
   V. Kralj-Igli\v{c} and A. Igli\v{c},
   {\em J. Phys. II France} {\bf 6}, 477 (1996). 

\bibitem{Rondelez}
   N. Cuvillier, M. Bonnier, F. Rondelez,
   D. Paranjape, M. Sastry and P. Ganguly,
   {\em Progr. Colloid Polym. Sci.} {\bf 105}, 118 (1997); 
   N. Cuvillier Ph.D thesis,
   University of Paris, France, 1997;
   N. Cuvillier and F. Rondelez,
   {\em Thin Solid Films} {\bf 329}, 19 (1998). 

\bibitem{prl}
   I. Borukhov, D. Andelman and H. Orland,
   {\em Phys. Rev. Lett.} {\bf 79}, 435 (1997). 

\bibitem{Kjellander}
   R. Kjellander, S. Mar\v{c}elja, R. M. Pashley and J. P. Quirk,
   {\em J. Chem. Phys.} {\bf 92}, 4399 (1990). 

\bibitem{Kekicheff}
  P. K\'ekicheff, S. Mar\v{c}elja, T. J. Senden and V. E. Shubin,
  {\em J. Chem. Phys.} {\bf 99}, 6098 (1993). 

\bibitem{HNC}
   R. Kjellander S. and Mar\v{c}elja,
   {\em J. Phys. Chem.} {\bf 90}, 1230 (1986); 
   R. Kjellander, T. \AA kesson, B. J\"onsson and S. Mar\v{c}elja,
   {\em J. Chem. Phys.} {\bf 97}, 1424 (1992). 

\bibitem{MSA}
   P. Attard, D. J. Mitchell and B. W. Ninham,
   {\em J. Chem. Phys.} {\bf 88}, 4987 (1988); 
                        {\bf 89}, 4358 (1988). 

\bibitem{Marcelo}
    J. Yu and M. Lozada-Cassou,
    {\em Phys. Rev. Lett.} {\bf 77}, 4019 (1996); 
    M. Lozada-Cassou, W. Olivares and B. Sulbar\'an,
    {\em Phys. Rev. E} {\bf 53}, 522 (1996). 

\bibitem{Hansen}
    E. Trizac and J. P. Hansen,
    {\em Phys. Rev. E} {\bf 56} 3137 (1997). 

\bibitem{Blankschtein}
    L. Lue, N. Zoeller and D. Blankschtein,
    {\em Langmuir} {\bf 15}, 3276 (1999). 

\bibitem{NetzOrland}
    R. R. Netz and H. Orland,
    {\em Europhys. Lett.} {\bf 45}, 726 (1999). 

\bibitem{Burak}
    Y. Burak and D. Andelman,
    {\em in preparation}.

\bibitem{Stern}
  O. Stern,
  {\em Z. Elektrochem.} {\bf 30}, 508 (1924). 

\bibitem{Coalson}
   Similar field theories leading to the Poisson-Boltzmann equation and
slight variations have been suggested earlier. See, e. g.,
  R. Podgornik and B. \v{Z}ek\v{s},
  {\em J. Chem. Soc. Faraday Trans.} {\bf 84}, 611 (1988); 
  R. Podgornik,
  {\em J. Phys. A: Math. Gen.} {\bf 23}, 275 (1990); 
%
  R. D. Coalson and A. Duncan,
  {\em J. Chem. Phys.} {\bf 97}, 5653 (1990); 
  R. D. Coalson, A. M. Walsh, A. Duncan and N. Ben-Tal,
  {\em J. Chem. Phys.} {\bf 102}, 4584 (1995). 

\bibitem{Wiegel}
   F. W. Wiegel,
   {\em Introduction to Path-Integral Methods in Physics and Polymer Science};
   World Scientific, Singapore (1986).

\bibitem{OrlandNegele}
      H. Orland and J. W. Negele,
       {\em Quantum Many-Particle Systems};
       Advanced Book Classics, Perseus Books (1988,1998).

\bibitem{Gurevich}
    Yu. Ya. Gurevich and Yu. I. Kharkats,
   {\em Dokl. Akad. Nauk SSSR} {\bf 229}, 367 (1976). 

\bibitem{Kornyshev}
     M. A. Vorotyntsev and A. A. Kornyshev,
   {\em Dokl. Akad. Nauk SSSR} {\bf 230}, 631 (1976); 
%
    A. A. Kornyshev and M. A. Vorotyntsev,
   {\em Electrochimica Acta} {\bf 26}, 303 (1981). 

\bibitem{GC}
    G. Gouy,
    {\em J. Phys. France} {\bf 9}, 457 (1910); 
    {\em Ann. Phys. (Paris)} {\bf 7}, 129 (1917); 
    D. L. Chapman,
   {\em Philos. Mag.} {\bf 25}, 475 (1913).

\bibitem{NetzOrland2}
    R. R. Netz and H. Orland,
    {\em to be published in Eur. Phys. J. E}.

\bibitem{pb-plus}
    I. Borukhov, {\em in preparation}.

\end{thebibliography}
\end{document}